\documentclass[a4paper,11pt]{article}
\usepackage{amsfonts}
\usepackage{amsmath}

\input{tcilatex}

\begin{document}

\author{Duoje Jia$^{\thanks{%
E-mail: jiadj@nwnu.edu.cn \ \ Supported by the Postdoctral Fellow Startup
Fund of NUNW(No. 5002--537).}\text{\ \ }\dagger }$ \\
$^{\ast }$Institute of Theoretical Physics, College of Physics\\
and Electronic Engineering, Northwest Normal \\
University,\textit{\ Lanzhou 730070, P.R. China}\\
$^{\dagger }$Interdisciplinary Center for Theoretical Study, \\
University of Science \& Technology of China, \\
Hefei, Aanhui 230026\textit{, P.R. China}}
\title{Dual Superconductivity from Yang-Mills Theory via Connection
Decomposition }
\maketitle
\date{}

\begin{abstract}
We derive an Abelian-Higgs-like action from $SU(2)$ Yang-Mills
theory via monopole-condensation assumption. Abelian projection as
well as chromo-'electric-magnetic' duality are naturally realized
by separating the small off-diagonal gluon part from diagonal
gluon field according to the order of inverse coupling
constant($1/g$). It is shown that Abelian dominance can follow
from infrared behavior of ranning coupling constant and the mass
generation of chromo-electric field as well as off-diagonal gluon
is due to the quantum fluctuation of orientation of Abelian
direction. Dual superconductivity of theory vacuum is confirmed by
deriving dual London equation for chromo-electronic field.

PACS number(s): 12.38.-t, 11.15.Tk, 12.38.Aw

\end{abstract}

\section{Introduction}

Although quantum chromodynamics(QCD) has been recognized to be the standard
theory of strong interactions, it fails to apply perturbatively in
low-energy regime where the effective coupling constant is expected to be
large and nonperturbative methods are needed. Till now, lattice QCD seems to
be the only effective approach to low-energy(nonperturbative) aspects of QCD
including the quark confinement\cite{Grainer}, except for various
phenomenological methods or semi-phenomenological methods (such as QCD sum
rules\cite{Shifman}). Recently, an appealing proposal\cite{FN} was made in
Yang-Mills(YM) theory by Faddeev and Niemi so as to separate collective
infrared variables from gauge field degrees of freedom via decomposing gauge
connection into an Abelian part, a unit color-vector $\mathbf{n}$ as well as
dual variables, which manifests the pure YM theory as an effective Abelian
theory with duality structure between chromo-electric and chromo-magnetic
field. This proposal is also called Faddeev-Niemi decomposition(FND). Owing
to absence of quark sources, this proposal was usually adopted in revealing
the knotted-vortex structure of QCD vacumm\cite{Faddeevknot} or the
gluonball spectrum\cite{glunball}, rather than the
dual-superconductor-oriented confining mechanism.

On the other hand, the close link of FND to the Abelian Projection(AP)
proposed by 't Hooft\cite{tHooftB455}, and the underlying dual structure
inherent in FND made it very suggestive and remarkable, and motivate ones to
inquire how FND bears on the dual-superconductor picture \cite%
{Numbu,tHooft76} of quark confinement as well as the supposed monopole
condensation. As shown by lattice simulations\cite{Suzuki,Suganuma} in the
maximal Abelian gauge(MAG), dual Abelian dynamics of QCD dominates in
infrared regime, where monopole degrees of freedom forms a condensate
responsible for the needed dual Meissner effect. Furthermore, the lattice
simulations on center vortices and monopoles (see \cite%
{Greensite03,Suganuma0407} for a review) revive the interests of
continuum-theory analysis \cite{Dzhunushaliev,Jdj,Gaete04} of
nonperturbative QCD.

In this Letter we present a calculation procedure for effective
Abelian-Higgs-like action from $SU(2)$ Yang-Mills theory. This action is
derived by reformulating YM theory via FND of the gluon field. In this
reformation, Abelian projection as well as chromo-'electric-magnetic'
duality are naturally realized by separating the small off-diagonal gluon
part from diagonal gluon field according to the order of inverse coupling
constant($1/g $). It is shown that Abelian dominance in confining regime of
low-energy QCD can follow from infrared behavior of running
coupling-constant. To second order of $1/g$, we confirm dual
superconductivity of theory vacuum by deriving dual London equation for
chromo-electric field.

\section{Connection decomposition and Abelian projection}

QCD has a crucial feature that it almost has no free parameter and
asymptotic freedom. If we ignore quark energy we left with pure YM
interaction energy in which only one parameter, namely, the coupling
constant($g$), is available. This can be a good approximation to QCD in
low-energy limit and can find its analogy in the quantum theory of
ultra-cold trapped atomic gas\cite{Dalfovo} where kinetic energy of the gas
system is, via Thomas-Fermi approximation, neglected in contrast with
inter-atomic interaction in zero-temperature limit. We note that this
analogy seems also work for weak-interaction case of them as we know that
quarks with high energy tend to freedom and the particles in gas are almost
free at high-temperature. This makes the link of FND to the AP for QCD
confinement, the dual structure of reformulated YM\ theory to
dual-superconductor vacuum of QCD physically relevant. Motivated by these
link, we study dual-superconductor picture of QCD from the viewpoint of
FND-based reformulation of YM theory.

We begin with $SU(2)$ YM theory where (gluon field) connection $\mathbf{A}%
_{\mu }=A_{\mu }^{a}\tau ^{a}$ ($\tau ^{a}=\sigma ^{a}/2,a=1,2,3$) describes
6 transverse UV degrees of freedom. We use\ inner product $\tau ^{a}\cdot
\tau ^{b}\equiv 2Tr(\tau ^{a}\tau ^{b})=\delta ^{ab}$, $\mathbf{A}\cdot
\mathbf{B}\equiv A^{a}B^{a}$, and across product$\ \mathbf{A}\times \mathbf{B%
}=-i[\mathbf{A},\mathbf{B}]$ for short. To parameterize $\mathbf{A}_{\mu }$
in terms of monopole variables, we invoke the infrared 'magnetic' variable $%
\mathbf{n}$(=$n^{a}\tau ^{a}$), an unit vector in internal(color) space\cite%
{Duan}. This vector naturally provides an preferred direction, breaking $%
SU(2) $ to $U(1)$ by leaving residual $U(1)$ symmetry (rotation around $%
\mathbf{n}$) intact, as required by AP.

Solving $\mathbf{A}_{\mu }$ from $D_{\mu }\mathbf{n}-\partial _{\mu }\mathbf{%
n}=g\mathbf{A}_{\mu }\times \mathbf{n}$, where $g$ is coupling constant, one
gets
\begin{equation}
\mathbf{A}_{\mu }=A_{\mu }\mathbf{n}+g^{-1}\partial _{\mu }\mathbf{n}\times
\mathbf{n+b}_{\mu }  \label{deD}
\end{equation}%
where $A_{\mu }\equiv \mathbf{A}_{\mu }\cdot \mathbf{n}$ transforms as an
Abelian connection($A_{\mu }\rightarrow A_{\mu }+\partial _{\mu }\alpha /g$%
), \ for $U(1)$ rotation $U(\alpha )=e^{i\alpha \overrightarrow{n}\cdot
\overrightarrow{\sigma }/2\text{ }}$(the rotation around the direction $%
\mathbf{n}$) and $\mathbf{b}_{\mu }=g^{-1}\mathbf{n}\times D_{\mu }(\mathbf{A%
}_{\mu })\mathbf{n}$ is $SU(2)$ covariant. Here, the first part $A_{\mu }%
\mathbf{n}$ in RHS of (\ref{deD}), being the diagonal part of gluon field,
corresponds to Abelian subgroup $H=U(1)$ while the second and third terms,
both of which are orthogonal to $\mathbf{n}$ and being off-diagonal gluon
parts(or non-Abelian components of gluon field), correspond to non-Abelian
group orbit $SU(2)/H$. We note that (\ref{deD}) can be true variable change%
\cite{Shabanov} if one takes $\mathbf{b}_{\mu }$ itself as a gauge vector
field and further imposes two constraints on\textbf{\ }$\mathbf{b}_{\mu }$.
This is necessary for getting marginal contribution to the final effective
action which we do not consider in this paper.

Note that the second term in RHS of (\ref{deD}) does not depend upon the
original degrees of freedom $\mathbf{A}_{\mu }$, which implies $\mathbf{A}%
_{\mu }$ may has intrinsic structure, that is, it may serves as monopoles in
some of its components. This idea is due to Duan's work on multi-monopoles%
\cite{Duan} and has applied to non-Abelian instantons due to defect\cite%
{Jinstant}. To find all relevant variables, we further decompose $\mathbf{b}%
_{\mu }$ in terms of $\mathbf{n}$. Observed that the internal orbit space $%
SU(2)/H$ can be spanned by basis $\partial _{\mu }\mathbf{n}$ and $\partial
_{\mu }\mathbf{n}\times \mathbf{n}$, one can re-parameterize $\mathbf{b}%
_{\mu }$ as
\begin{equation}
\mathbf{b}_{\mu }=g^{-1}\rho \partial _{\mu }\mathbf{n}+g^{-1}\sigma
\partial _{\mu }\mathbf{n}\times \mathbf{n.}  \label{furth}
\end{equation}%
Here, the scalar fields $\rho $ and $\sigma $ can be combined to define a
complex variable $\phi =\rho +i\sigma $. Substituting (\ref{furth}) into (%
\ref{deD}) we get the FND\cite{FN} for $SU(2)$ connection

\begin{equation}
\mathbf{A}_{\mu }=A_{\mu }\mathbf{n}+g^{-1}\partial _{\mu }\mathbf{n}\times
\mathbf{n}+g^{-1}\rho \partial _{\mu }\mathbf{n}+g^{-1}\sigma \partial _{\mu
}\mathbf{n}\times \mathbf{n,}  \label{deF}
\end{equation}%
in which $A_{\mu }$ (and $\mathbf{A}_{\mu }$) has dimension of [mass], $%
\mathbf{n}$, $\rho $ and $\sigma $ of unit.

We need to know the transformation role of all new variables under the
residual symmetry $U(\alpha )$. The transformation role of $\rho $ and $%
\sigma $ can be given by covariance of $\mathbf{b}_{\mu }$ under the
rotation $U(\alpha )$. Noticing that $[\partial _{\mu }\mathbf{n,}%
e^{-i\alpha \mathbf{n}}]=\alpha \partial _{\mu }\mathbf{n}\times \mathbf{n}$%
, $[\partial _{\mu }\mathbf{n\times n,}e^{-i\alpha \mathbf{n}}]=-\alpha
\partial _{\mu }\mathbf{n}$, one finds
\begin{eqnarray*}
\mathbf{b}_{\mu }^{U} &=&g^{-1}e^{i\alpha \mathbf{n}}(\rho \partial _{\mu }%
\mathbf{n}+\sigma \partial _{\mu }\mathbf{n}\times \mathbf{n)}e^{-i\alpha
\mathbf{n}}, \\
\ &=&g^{-1}(\rho -\alpha \sigma )\partial _{\mu }\mathbf{n}+g^{-1}(\sigma
+\alpha \rho )\partial _{\mu }\mathbf{n}\times \mathbf{n,}
\end{eqnarray*}%
which implies $\delta \rho =-\alpha \sigma \ $and $\delta \sigma =\alpha
\rho $, or%
\begin{equation*}
\delta (\rho +i\sigma )=i\alpha (\rho +i\sigma ).
\end{equation*}%
Thus, the complex variables $\phi $ indeed transforms as a charged scalar:%
\begin{equation*}
\phi \rightarrow \phi e^{i\alpha }.
\end{equation*}
However, it can be seen that the second term in RHS\ of (\ref{deF}) is not $%
U(1)$ covariant. In fact, this term has the form of non-Abelian monopole
potential\cite{FN}.

Since the connection $\mathbf{A}_{\mu }$ has $12$ field components while the
RHS of (\ref{deF}) has $8$ degrees of freedom, corresponding to 4 components
of $A_{\mu }$, $2$ independent components of $n^{a}$ and $2$ components $%
(\rho ,\sigma )$, the new variables ($A_{\mu }$,$n^{a}$,$\phi $) are still
short of $4$ degrees for variable change (\ref{deF}). If one would further
fixes $2$ longitudinal components of $U(1)$ connection $A_{\mu }$, the
resulted $6$ degrees of freedom of new variables corresponds to fully
gauge-fixed degrees(6 on-shell polarization components) of original gluon
field $\mathbf{A}_{\mu }$. Therefore, FND (\ref{deF}) with localized
variable $\mathbf{n}(x)$ serves as one example of partial gauge fixing used
in AP.

According to the original idea of 't Hooft\cite{tHooftB455}, Abelian
projection is realized by fixing the non-Abelian part of the gauge
ambiguity, breaking full gauge symmetry(that is $SU(2)$ here) into that of
maximal Abelian subgroup($U(1)$ here). The singularities in gauge condition
lead to difference between two group manifolds and were interpreted as
magnetic monopoles in the projected $U(1)$ gauge theory. Notice that (\ref%
{deF}) eliminates 4 degrees and it does not obey full gauge transformation
law, for instance, it fails to transform as connection under gauge rotation
around the direction different from $\mathbf{n}$, we know that (\ref{deF})
corresponds to the singular gauge in which $A_{\mu }\mathbf{n}$ is the
un-fixed diagonal variable and the other off-diagonal terms are non-Abelian
components whose degrees of freedom have been reduced.

\section{Chromo-'Electron-magnetic' duality and Abelian dominance}

Corresponding to gauge symmetry breaking $SU(2)\rightarrow U(1)$, we assume
the physical vacuum of infrared theory forms the monopole condensate,
sharing the residual symmetry of $U(1)$ rotation around direction $\mathbf{n}
$. To show Abelian dominance and dual structure of infrared YM theory, we
use the well-justified energy-dependence of effective coupling constant: $%
g_{s}^{2}\sim 1/B\log (Q^{2}/\Lambda ^{2})$, where \ $B=-\beta _{0}$ $>0$ is
the negative $\beta $-function of QCD\cite{Gross} at loop level. That means,
being converse limit of QCD asymptotic freedom, the effective coupling $%
g_{s} $ becomes sufficiently large in low-energy($Q$) limit (or infrared
limit). For simplicity, we use to $g$ denote the effective coupling
hereafter.

With (\ref{deF}), one finds the projection of non-Abelian gauge field $%
\mathbf{G}_{\mu \nu }$ along $\mathbf{n}$ to be $\ $%
\begin{equation}
\mathbf{G}_{\mu \nu }\cdot \mathbf{n=}F_{\mu \nu }+B_{\mu \nu }+g^{-1}%
\mathbf{n}\cdot (D_{\mu }\mathbf{n}\times D_{\nu }\mathbf{n}),  \label{gn}
\end{equation}%
where $F_{\mu \nu }\equiv \partial _{\mu }A_{v}-\partial _{v}A_{\mu }$ and
\begin{equation}
B_{\mu \nu }\equiv -g^{-1}\mathbf{n}\cdot (\partial _{\mu }\mathbf{n}\times
\partial _{\nu }\mathbf{n})  \label{h}
\end{equation}%
stands for the chromo-electric and chromo-magnetic field strengths,
respectively. One can identify magnetic potential $C_{\mu }$ by $B_{\mu \nu
}\equiv \partial _{\mu }C_{\nu }-\partial _{\nu }C_{\mu }$, where
parametrization of $C_{\mu }$ can not to be given in single-valued way. One
can calculate the magnetic charge $G_{m}$ by surface integral:
\begin{equation}
G_{m}=\int_{V^{(3)}}k_{0}d^{3}\sigma ^{0}  \label{2g}
\end{equation}%
where $k_{0}$ is the time component of the magnetic current
\begin{equation*}
k_{\mu }=\partial _{\nu }^{\ast }B_{\mu \nu },^{\ast }B_{\mu \nu }=\frac{%
\epsilon ^{\mu \nu \rho \lambda }}{2}B_{\rho \lambda }\text{.}
\end{equation*}%
With (\ref{h}) and (\ref{2g}), one can get%
\begin{eqnarray*}
G_{m} &=&\int_{V^{(3)}}\frac{\epsilon ^{0\nu \rho \lambda }}{2}\partial
_{\nu }B_{\rho \lambda }d^{3}x \\
&=&-\frac{1}{g}\int_{V^{(3)}}\frac{\epsilon ^{0\nu \rho \lambda }}{2}%
\partial _{\nu }\mathbf{n}\cdot (\partial _{\rho }\mathbf{n}\times \partial
_{\lambda }\mathbf{n})d^{3}x \\
&=&-\frac{4\pi }{g}\int_{V^{(3)}}\frac{\epsilon ^{\nu \rho \lambda }\epsilon
_{abc}}{8\pi }\partial _{\nu }n^{a}\partial _{\rho }n^{b}\partial _{\lambda
}n^{c}d^{3}x,
\end{eqnarray*}%
The computation of this integration gives\cite{Duan}

\begin{eqnarray}
G_{m} &=&-\frac{4\pi }{g}\int_{V^{(3)}}\delta (\Phi )J(\Phi /x)d^{3}x  \notag
\\
&=&-\sum_{i}\frac{4\pi w_{i}(\mathbf{n})}{g}  \label{gm}
\end{eqnarray}%
in which $\Phi =(\Phi ^{1},\Phi ^{2},\Phi ^{3})$ is defined as a vector
field along $\mathbf{n}$, i.e., $n^{a}=\Phi ^{\alpha }/\left\Vert \Phi
\right\Vert $ and $w_{i}(\mathbf{n})$ is the winding number(topological
charge) of $\Phi $ around its $i$-th singularity, with sign determined by
that of spacial Jacabian function $J(\Phi /x)$. Since $G_{m}$ stands for the
total magnetic charge in $V^{(3)}$, one can write (\ref{gm})
\begin{equation}
G_{m}=-\sum_{i}g^{(i)};g^{(i)}=\frac{4\pi w_{i}(\mathbf{n})}{g},  \label{gi}
\end{equation}%
where $g^{(i)}$ is the magnetic charge in $i$-th region $V^{(i)}$.

Let us parameterize $\mathbf{n}$ in terms of spherical coordinates as $%
\mathbf{n}_{0}=(\sin \gamma \cos \beta ,\sin \gamma \sin \beta ,\cos \gamma
) $. One has $C_{\mu }=g^{-1}(\cos \gamma \partial _{\mu }\beta +\partial
_{\mu }\alpha )$, which has a degree of freedom of $U(1)$ gauge
transformation ($C_{\mu }\rightarrow C_{\mu }+g^{-1}\partial _{\mu }\alpha $%
). This $U(1)$ symmetry is happened to hold simultaneously for Abelian part $%
A_{\mu }$. To see more specifically the link of FND\ to AP, let us take $%
\mathbf{n}_{0}$ along $\sigma ^{3}$ and write general gauge rotation as $%
U=e^{-\sigma ^{3}\alpha }e^{-\sigma ^{2}\gamma }e^{-\sigma ^{3}\beta }$.
Choosing a preferred direction $\mathbf{n}_{0}$ at each point implies we
keep the partial symmetry under rotation $e^{-\sigma ^{3}\alpha }$, with $%
e^{-\sigma ^{2}\gamma }e^{-\sigma ^{3}\beta }$ symmetry broken(with $\gamma $
and $\beta $ fixed). That means, $A_{\mu }=$ $\mathbf{A}_{\mu }\cdot \mathbf{%
n}_{0}$ and $C_{\mu }$ are both defined up to rotation $e^{-\sigma
^{3}\alpha \text{ }}$($\alpha \rightarrow \alpha +\alpha _{0}$). Therefore,
AP responds, in $SU(2)$\ case, to assigning specific direction $\mathbf{n}%
(x) $ at each spacetime point $x$ in FND (\ref{deF}).

To see Abelian dominance and duality in infrared YM theory specifically, we
take infinity limit of coupling $g$. First, we note that as a physical
field, the field exhibiting 'electron-magnetic' duality in color space
should be gauge-invariant and such a field tensor has ever be given by the
't Hooft \cite{tHooft},
\begin{eqnarray*}
f_{\mu \nu } &=&G_{\mu \nu }\cdot \mathbf{n}-g^{-1}\mathbf{n}\cdot (D_{\mu }%
\mathbf{n}\times D_{\nu }\mathbf{n}) \\
&=&F_{\mu \nu }+B_{\mu \nu }.
\end{eqnarray*}%
in which 'electric' and 'magnetic' field are put in the equal foots. The
total flux for the 't Hooft tensor $f_{\mu \nu }$ is
\begin{eqnarray}
\text{f\_Flux} &=&\int_{\Sigma }\frac{1}{2}(F_{\mu \nu }+B_{\mu \nu
})dx^{\mu }\wedge dx^{\nu }  \notag \\
&=&\doint\limits_{\partial \Sigma }A_{\mu }dx^{\mu }+\int_{V^{(\Sigma
)}}\partial _{\nu }^{\ast }B_{\mu \nu }d^{3}\sigma ^{\mu }  \notag \\
&=&\Phi _{e}(V^{(\Sigma )})+G_{m}(V^{(\Sigma )}),  \label{flux}
\end{eqnarray}%
where (\ref{2g}) was used. This duality is perfect and well established in
non-Abelian gauge theory with Higgs field, to which the internal direction
for AP is oriented to \cite{tHooft}.

In contrast, our reformulated theory appears as an Abelian gauge theory in
effective media of off-diagonal gluons, which correct the perfect Maxwell
theory with monopoles by a media-factor for 'magnetic' field $B_{\mu \nu }$.
This can be shown by calculating the non-Abelian gauge field $\mathbf{G}%
_{\mu \nu }$ in large-$g$ limit(IR limit). \ To the first order of $g^{-1}$,
one finds
\begin{equation*}
\mathbf{G}_{\mu \nu }\rightarrow \mathbf{n[}F_{\mu \nu }+(1-\rho ^{2}-\sigma
^{2}-2\sigma )B_{\mu \nu }]\text{,}
\end{equation*}%
which means the dominant part of gluon lines is distributed along Abelian
component. This transforms the standard YM theory into an Abelian gauge
theory with 'electric-magnetic' duality ($A_{\mu }\leftrightarrow C_{\mu }$)
\begin{eqnarray}
\mathfrak{L}_{dual} &=&\frac{1}{4}(F_{\mu \nu }+H_{\mu \nu })^{2}.
\label{dual} \\
H_{\mu \nu } &:&=Z(\phi )B_{\mu \nu },Z(\phi )=(1-\left\vert \phi
\right\vert ^{2}-2\func{Im}\phi ).  \label{HZ}
\end{eqnarray}%
We see that the infrared approximation of YM theory become the diagonal one
with $N-1=2$ types of Abelian charges, with magnetic charge dressed from the
off-diagonal variable $\phi $. When sources included, (\ref{dual}) is dual
to the QED with magnetic monopoles with electric charge vacuum-polarized and
the 'electric-magnetic' duality takes the form%
\begin{equation}
F_{\mu \nu }\leftrightarrow B_{\mu \nu };Z(\phi )\leftrightarrow 1/Z(\phi
),g\leftrightarrow 1/g.  \label{DD}
\end{equation}

From (\ref{dual}) and (\ref{HZ}) we see that due to the contribution of
off-diagonal gluons the duality otherwise manifested by 't Hooft tensor $%
f_{\mu \nu }$ in infrared regime was replaced in our reformulated theory by
that with effective magnetic media-factor $Z(\phi )$. We also see that the
duality between the chromo-electric and chromo-magnetic field(EM) strongly
depends upon the Abelian dominance in the sense that the duality becomes
exact as Abelian gluon part is getting dominant.

Breaking of $SU(2)\rightarrow U(1)$, or fixing of the direction of $\mathbf{n%
}$ (quantum operator) at each point $x$ makes it acquire nonvanishing vacuum
expectation value(vev.), $\langle n^{a}(x)\rangle =n^{a}(x)$ (c-number
field), and
\begin{equation}
\langle \partial ^{\mu }n^{a}(x)\partial _{\nu }n^{a}(x)\rangle =\delta
_{\nu }^{\mu }\langle (\partial n^{a})^{2}\rangle =-\delta _{\nu }^{\mu
}m^{2}\   \label{vev}
\end{equation}%
in which $m$ is a mass scale and the minus sign comes from the fact $%
\partial _{\mu }\mathbf{n}$ is space-like for our static case. Here, $\delta
_{\nu }^{\mu }$ arises from the requirement of Lorentz invariance of vev.
and $m^{2}$ is due to that $\partial _{\mu }n^{a}$ may has a normalized
factor (quantum fluctuation of direction of $\mathbf{n}$) and it has
dimension of [Mass]. We can interpret such a behavior as the particles
associated with $\mathbf{n}$-field undergo condensation in theory\ vacuum,
or in other words, the $S^{2}$ symmetry (rotation of $\mathbf{n}$%
-orientation) of the original theory was broken by the QCD\ vacuum. One then
has
\begin{eqnarray*}
\langle C_{\mu }^{2}\rangle &=&g^{-2}\langle (\partial \mathbf{n}%
)^{2}\rangle =-g^{-2}m^{2} \\
\langle B_{\mu \nu }\rangle &=&0
\end{eqnarray*}%
since $B$ is anti-symmetric. Furthermore, the vev.'s of all field components
with Lorentz indices or color indices explicitly, such as $\langle A_{\mu
}\rangle ,\langle C_{\mu }\rangle $ and $\langle \partial _{\mu
}n^{a}\rangle $, vanish since these vev.'s become physical in condensate and
thereby they are Lorentz invariant and gauge invariant.

Taking these consideration into account, one finds%
\begin{equation}
\langle \mathfrak{L}_{dual}\rangle =-\frac{1}{4}F_{\mu \nu }^{2}-\frac{1}{4}%
\lambda Z(\phi )^{2},  \label{FZ}
\end{equation}%
where $\lambda \equiv \langle B_{\mu \nu }^{2}\rangle $ is positive scale
and with dimension of 4. One can see that the duality survives in
purely-diagonal gluon dynamics (\ref{FZ}) but dual-superconductivity does
not since this dynamics is short of crucial ingredient, kinetic term of $%
\phi $ field, which makes (\ref{FZ}) into an effective superconductor
model---Abelian Higgs model. In the nest section, we will find such a
ingredient could present when we include the contribution of off-diagonal
gluon.

\subsection{Abelian Higgs action in terms of collective dual variables}

As a relativistic generalization of effective superconductor model--the
Ginzburg-Landau model, the Abelian Higgs model has long been proposed to
describe the confining phase of QCD, in which the string-like singularities
provide the confining forces between field sources \cite{Numbu,tHooft76}. In
addition to the pure duality analysis via complete Abelian dominance, given
in last section, we here include the off-diagonal gluon contribution to the
order of $g^{-2}$.

With (\ref{deF}), one finds
\begin{eqnarray}
\mathbf{G}_{\mu \nu } &=&\mathbf{n[}F_{\mu \nu }+(1-\rho ^{2}-\sigma
^{2}-2\sigma )B_{\mu \nu }]+  \notag \\
&&(g^{-1}\nabla _{\mu }\rho +2A_{\mu })\partial _{\nu }\mathbf{n}%
-(g^{-1}\nabla _{\nu }\rho +2A_{\nu })\partial _{\mu }\mathbf{n}  \notag \\
&&+g^{-1}\nabla _{\mu }\sigma \partial _{\nu }\mathbf{n}\times \mathbf{n}%
\mathbb{-}g^{-1}\nabla _{\nu }\sigma \partial _{\mu }\mathbf{n}\times
\mathbf{n},  \label{G}
\end{eqnarray}%
where $n_{\mu \nu }=\delta _{\mu \nu }(\partial _{\rho }\mathbf{n}%
)^{2}-\partial _{\mu }\mathbf{n}\cdot \partial _{\nu }\mathbf{n}$, $\nabla
_{\mu }\rho =\partial _{\mu }\rho +gA_{\mu }\sigma $ and $\nabla _{\mu
}\sigma =\partial _{\mu }\sigma -gA_{\mu }\rho $. With (\ref{G}), one gets
\begin{eqnarray}
\mathfrak{L}_{dual} &=&-\frac{1}{4}\{F_{\mu \nu }^{2}+(1-\rho ^{2}-\sigma
^{2}-2\sigma )^{2}B_{\mu \nu }^{2}+  \notag \\
&&2(1-\rho ^{2}-\sigma ^{2}-2\sigma )F_{\mu \nu }B^{\mu \nu }  \notag \\
&&+\frac{2n_{\mu \nu }}{g^{2}}(\nabla ^{\mu }\rho +2gA^{\mu })(\nabla ^{\nu
}\rho +2gA^{\nu })  \notag \\
&&+\frac{2n_{\mu \nu }}{g^{2}}\nabla ^{\mu }\sigma \nabla ^{\nu }\sigma \}.
\label{L0}
\end{eqnarray}%
It is useful to define a $U(1)$ covariant derivative
\begin{eqnarray*}
\nabla _{\mu }\phi &=&\nabla _{\mu }\rho +i\nabla _{\mu }\sigma \\
&=&(\partial _{\mu }-igA_{\mu })\phi \text{.}
\end{eqnarray*}%
Here, we look $\phi $ as a charged field strongly coupled with 'electric'
field $A_{\mu }$ by strength $g$. Then, by averaging (\ref{L0}) over $%
\mathbf{n}$ with
\begin{equation}
\langle n_{\nu }^{\mu }\rangle =\delta _{\nu }^{\mu }\langle (\partial
n^{a})^{2}\rangle =-\delta _{\nu }^{\mu }m^{2},  \label{msq}
\end{equation}%
one can get the effective Lagrangian
\begin{eqnarray}
\mathfrak{L}^{eff} &=&-\frac{1}{4}F_{\mu \nu }^{2}+\frac{m^{2}}{2g^{2}}%
|\nabla _{\mu }\phi |^{2}+2m^{2}A_{\mu }^{2}  \notag \\
&&+\frac{2m^{2}}{g}A^{\mu }\func{Re}(\nabla _{\mu }\phi )-V(\phi ),
\label{AH1}
\end{eqnarray}%
which is Abelian-Higgs like model and
\begin{equation*}
V(\phi )=\frac{\lambda }{4}(|\phi |^{2}-1+2\func{Im}\phi )^{2}.
\end{equation*}

Here, $\lambda $ can be shown to be%
\begin{equation}
\lambda =\frac{2m^{4}}{g^{2}}.  \label{lum}
\end{equation}%
In fact, from (\ref{vev}) and invariance of the vev., one has%
\begin{eqnarray*}
\langle B_{\mu \nu }^{2}\rangle &=&g^{-2}\epsilon _{abc}\epsilon
_{mkl}\langle n^{a}\rangle \langle n^{m}\rangle \langle \partial _{\mu
}n^{b}\partial ^{\mu }n^{k}\rangle \langle \partial _{\nu }n^{c}\partial
^{\nu }n^{l}\rangle \\
&=&g^{-2}\epsilon _{abc}\epsilon _{mkl}n^{a}n^{m}\delta ^{bk}\langle
(\partial \mathbf{n})^{2}\rangle \delta ^{cl}\langle (\partial \mathbf{n}%
)^{2}\rangle \\
&=&g^{-2}\epsilon _{abc}\epsilon _{mbc}n^{a}n^{m}m^{4} \\
&=&2!g^{-2}m^{4}.
\end{eqnarray*}%
The effective Lagrangian becomes
\begin{eqnarray}
\mathfrak{L}^{eff} &=&-\frac{1}{4}F_{\mu \nu }^{2}+\frac{m^{2}}{2g^{2}}%
|\nabla _{\mu }\phi |^{2}-V(\phi )  \notag \\
&&+2m^{2}(1+\func{Im}\phi )A_{\mu }^{2}+\frac{2m^{2}}{g}A^{\mu }\func{Re}%
(\partial _{\mu }\phi ),  \label{AH3}
\end{eqnarray}

We see here that, as a consequences of $\mathbf{n}$-field condensation, not
only do the off-diagonal gluons gain mass but also has Abelian gluon field $%
A_{\mu }$ acquired a mass $\sim m$. We obtain massive gluon in our effective
theory(\ref{AH3}) without invoking the Higgs-like spontaneous
symmetry-breaking(SSB) mechanism as done in superconductor\ theory or dual
superconductor picture for confinement. In contrast, it is due to SSB of a
color direction field $\mathbf{n}(x)$ in the nontrivial QCD vacuum, provided
that magnetic-charges Bose condensed. The lattice simulation\cite%
{Suganuma,Giacomo} has confirmed\ magnetic-charge condensation in MAG. For $%
SU(2)$ theory, it is easy to see that change of variables (\ref{deF})
corresponds to MAG, since $U(1)$ rotation around $\mathbf{n}$ is maximal
Abelian subgroup of $SU(2)$. Therefore, our effective theory(\ref{AH3}) has
the key feature of dual superconductor and in this theory desired mass
generates from the off-gluon field $\mathbf{n}$. In fact, our calculation
from (\ref{msq}) to (\ref{AH3}) shows that such a mass generation arises
from quantum fluctuation of orientation $\mathbf{n}$, see (\ref{msq}) and (%
\ref{lum}).

We note that the marginal terms have not included in (\ref{AH1}) since new
variables (\ref{deF}) are on-shell degrees of freedom. The marginal-term
inclusion can be done by using off-shell field decomposition\cite%
{Faddeevknot} and then calculating the effective action through quantum
partition functional $Z\sim \int [dn^{a}]e^{-iS}$ . Toward the leading
infrared term, however, our model is sufficient for the effective
description of low-energy YM theory(\ref{AH1}).

\subsection{Dual London equation}

\bigskip To see the relation between the AP and mass generation of Abelian
gluon field, we need dual Meissner effect as a possible signal of the
monopole condensation. Here, we show that the effective model of QCD can
yield, in low-energy limit, a dual London equation for Abelized fields.
Varying (\ref{AH3}) gives
\begin{eqnarray}
\partial _{\mu }F^{\mu \nu } &=&\frac{m^{2}}{2g^{2}}[i\phi ^{\ast }%
\overleftrightarrow{\partial ^{\nu }}\phi -2\func{Re}\partial ^{\nu }\phi
]-m^{2}|\phi +2|^{2}A^{\nu }  \notag \\
\nabla _{\mu }\nabla ^{\mu }\phi &=&-\frac{\partial V(\phi )}{\partial \phi
^{\ast }}-\frac{m^{2}}{g^{2}}\partial ^{\mu }A_{\mu }+im^{2}A_{\mu }^{2}%
\text{, and c.c.,}  \label{EL}
\end{eqnarray}%
where
\begin{equation*}
\frac{\partial V(\phi )}{\partial \phi ^{\ast }}=\frac{m^{2}}{g^{2}}[|\phi
|^{2}-1-2\func{Im}\phi ](\phi +i).
\end{equation*}%
We see that the chromo-electric field $A_{\mu }$ strongly coupled with the
charged scalar field $\phi $ with coupling $g$ while $\phi $ is weakly
coupled to itself in the effective dynamics.

Taking the $g\rightarrow \infty $ limit in (\ref{EL}) and using Lorenz gauge
for $A^{\mu }$, we find
\begin{eqnarray}
\phi &\approx &\phi _{0}=-im^{2}/g^{2}\text{, and c.c.,}  \notag \\
\partial _{\mu }F^{\mu \nu } &=&j^{\nu }=-m_{V}^{2}A^{\nu },  \label{London}
\end{eqnarray}%
in which the second equation (\ref{London}) takes the form of London's
equation. Here%
\begin{equation*}
m_{V}=m(4+m^{4}/g^{4})^{1/2}\approx 2m,
\end{equation*}%
is the mass scale responsible for dual Meissner effect and its inverse $%
\lambda _{L}=1/m_{V}$ determines the transverse dimensions of the
chromo-electric field $A_{\mu }$ penetrating into the vacuum condensate. As
in superconductor, (\ref{London}) implies that chromo-electric field decays
as$\ $%
\begin{equation*}
A_{\mu }(d)=A_{\mu }(0)\exp (-d/\lambda _{L})
\end{equation*}%
\ as they depart from the singular vortex tube(string), where $d$ stands for
the distance away from string. This is consistent with the dual
superconductor picture\cite{Numbu}.

We note that similar argument for deriving an equation (\ref{London}) was
also given by Dzhunushaliev\cite{Dzhunushaliev} via ordered Abelian
components assumption and AP. It should be pointed out that in his approach
the 'electron-magnetic' duality needed (\ref{DD}) for dual-superconductivity
is not exhibited explicitly and the link of the generation of the vector
field mass $m_{V}$ with the magnetic charge condensation is not clarified.
We also note that the uniform assumption for scalar field $\phi $ in Ref.%
\cite{Dzhunushaliev} only follows in infrared limit, or large-$g$ limit.

In conclusion, we calculated an effective Abelian-Higgs-like action based on
the Faddeev-Niemi decomposition of $SU(2)$ gauge field. Abelian projectional
and chromo-'electric-magnetic' duality are realized via the decomposition in
which gluon fields are divided into $U(1)$\ diagonal part and small
non-Abelian off-diagonal parts with order of inverse coupling-constant($1/g$%
). We have shown that Abelian dominance can follow from infrared behavior of
ranning coupling constant and the mass generation for chromo-electric field
as well as off-diagonal gluon can arise from quantum fluctuation of
orientation of the unit iso-vector $\mathbf{n}$. Furthermore, we have
derived a dual London equation for chromo-electric field. This enhances the
dual superconductor picture as the possible mechanism of quark confinement.

D. Jia thanks X-J Wang and J-X Lu for numerous discussion, and M-L. Yan for
valuable suggestions.


\begin{thebibliography}{99}
\bibitem{Grainer} W. Grainer and A. Sch\"{a}fer, Quantum Chromodynamics,
Springer-Verlag, (Berlin Heidelberg, 1994).

\bibitem{Shifman} M. A. Shifman et al.,  Nucl. Phys. B\textbf{147,}
385(1979); Nucl. Phys. B\textbf{147,} 448 (1979).

\bibitem{FN} L. D. Faddeev and A. J. Niemi,  Phys. Rev. Lett. \textbf{82},
1624 (1999).

\bibitem{Faddeevknot} L. D. Faddeev and A. J. Niemi, Phys. Lett. B\textbf{449%
}, 214(1999);Phys. Lett. B\textbf{464}, 90(1999).

\bibitem{glunball} Y. M. Cho,  Phys. Rev. Lett. \textbf{87}, 252001(2001).

\bibitem{tHooftB455} G. 't Hooft,  Nucl. Phys. B[\textbf{FS3}]\textbf{190},
455 (1981).

\bibitem{Numbu} Y. Numbu, Phys. Rev. D\textbf{10}, 4262 (1974); S.
Mandelstam, Phys. Rep. C\textbf{23}, 245 (1976).

\bibitem{tHooft76} G. 't Hooft, Nucl. Phys. B\textbf{138},1 (1978); A. M.
Polyakov, Nucl. Phys. B\textbf{120}. 429 (1977).

\bibitem{Suzuki} T. Suzuki et al., Phys. Rev. D\textbf{42}, 4257 (1990); J.
Stack et al., Phys. Rev. D\textbf{50}, 3399 (1994); G. Bali et al., Phys.
Rev. D\textbf{54}, 2863 (1996).

\bibitem{Suganuma} H. Suganuma et al.,  Phys. Rev. D\textbf{60}, 77501
(1999); Nucl. Phys. B\textbf{548} 365 (1999); Nucl. Phys. B\textbf{574} 70
(2000).

\bibitem{Greensite03} J. Greensite, Prog. Part. Nucl. Phys. \textbf{51}, 1
(2003).

\bibitem{Suganuma0407} H. Suganuma et al., hep-ph/0407123; hep-lat/0407012.

\bibitem{Gross} D. Gross and F. Wilczek, Phys. Rev. Lett. \textbf{30}, 1343
(1973); D. Politzer, Phys. Rev. Lett. \textbf{30, }1346 (1973).

\bibitem{tHooft} G. 't Hooft, Nucl. Phys. B\textbf{79}, 276 (1974).

\bibitem{Dzhunushaliev} V. Dzhunushaliev, Phys. Rev. D\textbf{65}, 125007
(2002).

\bibitem{Jdj} D. Jia  and X. G. Li, High Energ. Phys. and Nucl. Phys.
\textbf{4}, 293 (2003).

\bibitem{Gaete04} P. Gaete and E. I. Guendelman, Phys. Lett. B \textbf{593},
151 (2004).

\bibitem{Dalfovo} F. Dalfovo, S. Giorgini, et al., Rev. Mod. Phys. \textbf{71%
},\textbf{\ }463 (1999).

\bibitem{Duan} Y. S. Duan and  M. L. Ge,  Sci. Sin. \textbf{11}, 1072
(1979); Y. M. Cho,   Phys. Rev. D\textbf{21}, 1080 (1980); Li S et al.,
Phys. Lett. B\textbf{487}, 201 (2000). hep-th/9911132

\bibitem{Shabanov} S.\ V.\ Shabanov, Phys. Lett. B\textbf{463}, 263 (1999).

\bibitem{Jinstant} D. Jia and Y. S. Duan, Mod. Phys. lett. A \textbf{16},
1863 (2001).

\bibitem{Giacomo} A. Di. Giacomo, B. Lucini et al., Phys. Rev. D 61, 034503
(2000) [arXiv:hep-lat/9906024];Phys. Rev. D 61, 034504 (2000)
[arXiv:hep-lat/9906025];J. M. Carmona, M. D'Elia et al., Phys. Rev. D 64
,114507 (2001)[arXiv:hep-lat/0103005].
\end{thebibliography}
\end{document}